\documentclass[12pt,a4paper]{article} 
\usepackage{jheppub}

\title{BSM Physics: What the Higgs Can Tell Us}

\author[a]{Matthew S.~Brown}
\author[a]{Daniele Barducci}
\author[a,b]{Alexander Belyaev}
\author[c]{Stefania de Curtis}
\author[a]{Stefano Moretti}
\author[d]{Giovanni M.~Pruna}
\author[e]{Alexander Pukhov}

\affiliation[a]{Physics and Astronomy, University of Southampton, Highfield Campus, Southampton, SO17 1BJ, UK}

\affiliation[b]{Particle Physics Department, Rutherford Appleton Laboratory, Chilton, Didcot, Oxon OX11 0QX, UK}

\affiliation[c]{INFN, Sezione di Firenze, Via G. Sansone 1, 50019 Sesto Fiorentino, Italy}

\affiliation[d]{TU Dresden, Institut fu\"r Kern- und Teilchenphysik, Zellescher Weg 19, D-01069 Dresden, Germany}

\affiliation[e]{Skobeltsyn Institute of Nuclear Physics, Moscow State University, Moscow 119992, Russia}

\emailAdd{M.S.Brown@soton.ac.uk}
\emailAdd{A.Belyaev@soton.ac.uk}
\emailAdd{S.Moretti@soton.ac.uk}
\emailAdd{pukhov@lapth.cnrs.fr}
\emailAdd{decurtis@fi.infn.it}
\emailAdd{db3e11@soton.ac.uk}
\emailAdd{Giovanni\_Marco.Pruna@tu-dresden.de}

\abstract{This discovery of the Higgs boson last year has created new possibilities for testing candidate theories for explaining physics beyond the Standard Model.
Here we explain the ways in which new physics can leave its marks in the experimental Higgs data, and how we can use the data to constrain and compare different models. In this proceedings paper we use two models, Minimal Universal Extra Dimensions and the 4D Composite Higgs model, as examples to demonstrate the technique.}

\notoc
\begin{document}

\maketitle

\section{Introduction}

In July last year, the ATLAS and CMS collaborations jointly announced the discovery of a new particle of nature \cite{Aad:2012tfa, Chatrchyan:2012ufa} at the Large Hadron Collider (LHC). 
Since then, the challenge has been to measure the properties of the new resonance, and its couplings to other particles, to determine whether it is the long sought-after Higgs boson of the Standard Model (SM), or something more exotic. 
This opens the door to new techniques for testing theories that seek to go Beyond the Standard Model (BSM): such theories must reproduce the measured pattern of Higgs couplings.

In this paper we first outline the procedure for interpreting measurements of Higgs couplings and comparing them with predictions from a general BSM theory. 
In the two following sections, as examples, we show work we have done on two different models: Minimal Universal Extra Dimensions and the 4D Composite Higgs Model and show how the Higgs data constrain these models. 
Finally we state our conclusions.

\section{Higgs production and decay}
Studies involving a light Higgs (i.e., below the $WW$ theshold) generally work within the narrow width approximation (although see Ref.~\cite{Kauer:2012hd} for criticism of this approach). 
This means that the Higgs is assumed to be produced on-shell. The main mechanisms for SM Higgs production are gluon-gluon fusion ($ggH$), vector boson fusion (VBF), Higgs-strahlung ($VH$, where $V=W,Z$) and associated top production ($ttH$). 
For the SM Higgs, the dominant production process for the LHC is gluon fusion via a triangle loop of quarks, with the high LHC gluon luminosity making up for the loop suppression, together with a large one-body phase space. 
The Higgs can then decay directly (i.e. via a tree-level coupling) to any pair of massive particles, and it can decay to gluons and photons at the one-loop level. 
Despite the branching ratio to two photons being extremely small ($0.23 \%$ in the SM), it is a very sensitive channel due to the low irreducible background.

BSM physics can lead to a modification of the tree-level couplings of the Higgs to SM particles. 
New physics can also have an effect via new particles entering the loops in the $gg\to H$ production and/or the $H\to\gamma\gamma$ decay.

In the narrow width approximation, then, the number of Higgs signal events one would expect to be produced by channel $P$, initially decaying into channel $X$, is given by
\begin{equation}
    N_{PX}^c = \varepsilon_{PX}^c \times \sigma_{P} \times \text{BR}_X \times L,
\end{equation}
where $\sigma_{P}$ is the $pp\to HY$ Higgs production cross-section ($Y$ denoting other particles that might have been produced in association with the Higgs) in channel $P$, $\text{BR}_X$ is the Higgs branching ratio to the final state $X$, and $L$ is the integrated luminosity. 
The ATLAS and CMS collaborations split the events into different categories (denoted here by the superscript $c$'s), using sequences of cuts. 
The efficiency factor $\varepsilon_{PX}^c$ takes into account that only a fraction of the actual events will be accepted into a particular category $c$.

Unfortunately, the experimental collaborations do not make available the efficiency factors for every category considered, and they generally combine the different event categories for a particular decay channel into a combined value $\mu_X$. 
In the work presented in this paper we made the simplifying assumption that for each decay channel $X$ one production process $P$ dominates, thus
\begin{equation}
    \mu_X \approx \frac{\sigma_P \times \text{BR}_X}{\sigma_P^\text{SM} \times \text{BR}_X^\text{SM}} = \frac{\sigma_P}{\sigma_P^\text{SM}} \times \frac{\Gamma_X}{\Gamma_X^\text{SM}} \times \frac{\Gamma_\text{tot}^\text{SM}}{\Gamma_\text{tot}}
\end{equation}
(note that the unknown efficiency factors cancel). 
We assumed that gluon-gluon fusion dominates after cuts for all channels except for $H\to \bar{b}b$, where Higgs-strahlung dominates. 
If we denote the amplitude-level couplings of the Higgs to SM particles in SM units by $\kappa$, then the above equation can be written as
\begin{equation}
    \mu_X = \kappa_P^2  \kappa_X^2 / \kappa_H^2
\end{equation}
with
\begin{equation}
    \kappa_H^2 = \sum_X\left(\kappa_X^2 \times \text{BR}_X^\text{SM}\right).
\end{equation}
For the $H\to \bar{b}b$ channel $\kappa_P = \kappa_V$ (assuming that $W$ and $Z$ have their couplings enhanced by the same amount, relative to the SM values), and for all other decay channels we take $\kappa_P = \kappa_g$. 

Under the above assumptions, the process of testing a model simplifies to calculating the coupling enhancements in the model, forming the $\mu_X$ predictions and comparing these values to those found in experiment. 
To illustrate the process we present two example BSM theories in the following two sections and compare each of them with the Higgs data.

\section{Minimal Universal Extra Dimensions}

The work discussed in this section was done by G.~B\'elanger, A.~Belyaev, M.S.~Brown, M.~Kakizaki and A.~Pukhov, and it is published in Ref.~\cite{Belanger:2012mc}. 
The concept of Universal Extra Dimensions (UED) was proposed by Appelquist et al \cite{PhysRevD.64.035002} following previous suggestions of millimetre-scale extra dimensions \cite{ArkaniHamed:1998rs, antoniadis1998new} and warped extra dimensions \cite{PhysRevLett.83.3370}. 
In UED models, excitations of all fields are allowed to propagate in the extra dimension(s). 
Minimal Universal Extra Dimensions (MUED) proposes a single extra dimension, compactified on an $S^1/\mathbb{Z}_2$ ``orbifold'' in order to achieve chiral SM fermions. 
The field content is the same as the SM except that all fields are 5D. 
As an effective 4D theory, this is expressed as each 5D field being represented by a tower of progressively more-massive 4D fields (called Kaluza-Klein (KK) modes).
The zero mode of each 5D field furnishes the SM particle spectrum.

At tree level, the particle spectrum is incredible simple, with the masses of the KK particles at KK level $n$ being simply related to the associated SM mass $m_0$ via
\begin{equation}
    m_n = \sqrt{n^2/R^2 + m_0^2}.
\end{equation}
If this were the whole story, the KK particles would all be stable (there would be no phase space for decays to occur). 
In this respect, radiative corrections play a vital role in that they break the degeneracy of the masses and dictate which decays become allowed and forbidden. 
These mass corrections are calculated in Ref.~\cite{Cheng:2002iz}.

There is only one new parameter in this theory (beyond the SM parameters): the compactification radius $R$. 
Strictly speaking, the theory is non-renormalisable so one must introduce an ultraviolet cutoff $\Lambda$ as well. However, the observables of interest in this paper are only weakly sensitive to $\Lambda$ (which we take to be $\Lambda=20 R^{-1}$).

All of the tree-level Higgs couplings in MUED are identical to the SM values, so the only contribution to $\mu_X$ from new physics come from $\kappa_g$ and $\kappa_\gamma$. 
At one-loop level the $gg\to H$ amplitude enhancement is
\begin{equation}
\kappa_g = \frac{F_{ggH}^{\text{SM}} + \sum_{n=1}^N F_{ggH}^{(n)}}{F_{ggH}^{\text{SM}}}
\label{eq:gghmaster}
\end{equation}
where the sum is taken over the KK number $n$.
In the SM there would be contributions from each quark flavour $q$  in
the loop, such that  $F_{ggH}^{\text{SM}} = \sum_{q} f_{F}(m_q)$ where $f_F$ the standard fermion loop function (e.g. see Ref.~\cite{Djouadi:2005gi}).
The contribution from KK quarks at the $n$th KK level (there are two KK quarks at each level for each SM quark $q$) is
\begin{equation}
F_{ggH}^{(n)} = \sum_{q}\sin(2a_{q}^{(n)})
\left(\frac{m_q}{m_{q,1}^{(n)}} f_{F}(m_{q,1}^{(n)}) +
\frac{m_q}{m_{q,2}^{(n)}} f_{F}(m_{q,2}^{(n)}) \right)
\label{eq:fgghfull}
\end{equation}
where $m_{q,1}^{(n)}$ and $m_{q,2}^{(n)}$ denote the KK quark masses and
$a_{q}^{(n)}$ denote the mixing angles required to diagonalise the KK quark
mass matrices.

Similarly, $\kappa_\gamma$ is given by
\begin{equation} 
\kappa_\gamma = \frac{F_{H\gamma\gamma}^{\text{SM}} +
\sum_{n=1}^{N}F_{H\gamma\gamma}^{(n)}}{F_{H\gamma\gamma}^{\text{SM}}}
\end{equation}
with
\begin{equation}
F_{H\gamma\gamma}^{\text{SM}} = f_{V}(m_{W}) +
\sum_{f}n_{c}Q_{f}^{2}f_{F}(m_{f}).
\label{eq:FhAASM}
\end{equation}
The sum is taken over all SM fermions $f$, each with charge
$Q_{f}e$, setting $n_{c}$ to 3 for quarks and 1 for leptons.
The fermion loop function $f_{F}$ is the same as for the $gg\to H$ case and $f_V$ is the standard vector loop function \cite{Djouadi:2005gi} (representing the $W^\pm$ and related Goldstone and ghost contributions).
At the $n$th KK level the amplitude receives contributions from KK charged
fermions (two KK partners for each SM fermion) and the KK $W_{n}^\pm$ vector
boson. There is also a contribution from the charged scalar $a_n^\pm$ that
is not present at the SM level, so
\begin{equation}
	F_{H\gamma\gamma}^{(n)} = f_F^{(n)} + f_V^{(n)} + f_S^{(n)}.
\label{eq:fhgg:KK}
\end{equation}
The fermion contribution is the same as the quark contribution \eqref{eq:fgghfull} was for the Higgs production amplitude, but with an additional colour and charge factor of $n_c Q_f^2$ for each fermion flavour $f$.
The KK vector contribution is $f_V^{(n)} = \frac{m_W^2}{m_{W,n}^2}f_V(m_{W,n})$, and the scalar contribution is given by
\begin{align}
  f_{S}^{(n)}(m_{a,n},m_{W,n}) &=
  \left[\frac{2m_{W}^{2}}{m_{W,n}^{2}}\left(1-\frac{2m_{a,n}^{2}}{m_{H}^{2}}
  \right) - 2\right] \left[1-\frac{4m_{a,n}^2}{m_H^2}c_{0}(m_{a,n}) \right].
\label{eq:fS}
\end{align}

\begin{figure}[tb]
\begin{center}
\includegraphics[width=0.5\linewidth]{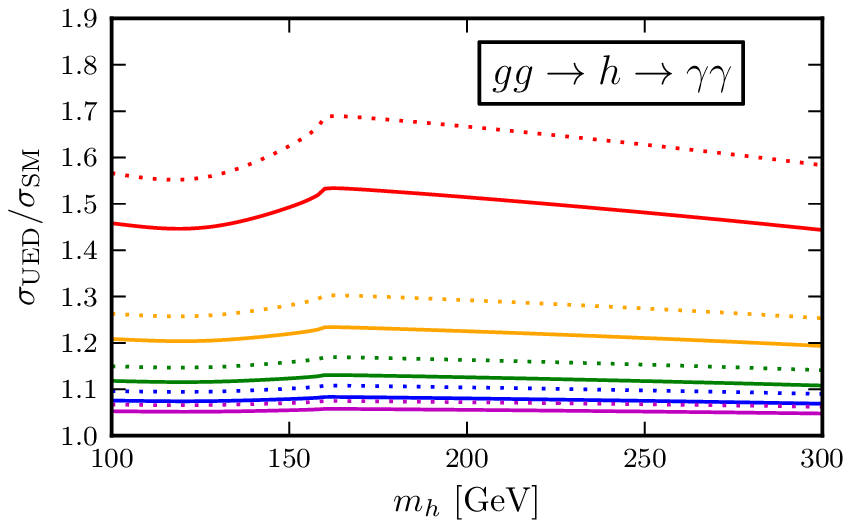}%
\includegraphics[width=0.5\linewidth]{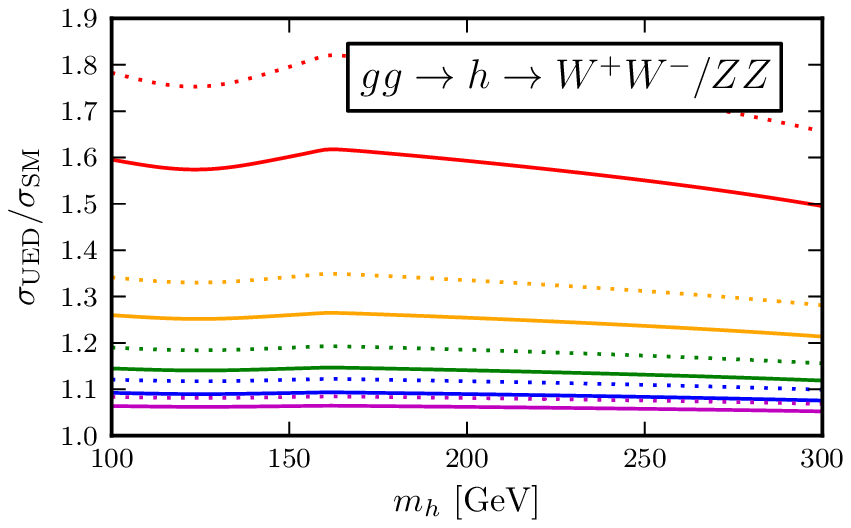}
\end{center}
\caption{Signal enhancement factors for the $H\to \gamma\gamma$, $H\to WW\to2\ell 2\nu$ and $H\to ZZ \to 4\ell$ channels for varying values of the compactification scale $R^{-1}$. From top to bottom, $R^{-1}=500$, 750, 1000, 1250 and $1500~\text{GeV}$. Results are shown with (solid lines) and without (dashed lines) radiative KK mass corrections.}
\label{fig:muedmu}
\end{figure}

\begin{figure}[tb]
\begin{center}
\includegraphics[width=0.6\linewidth]{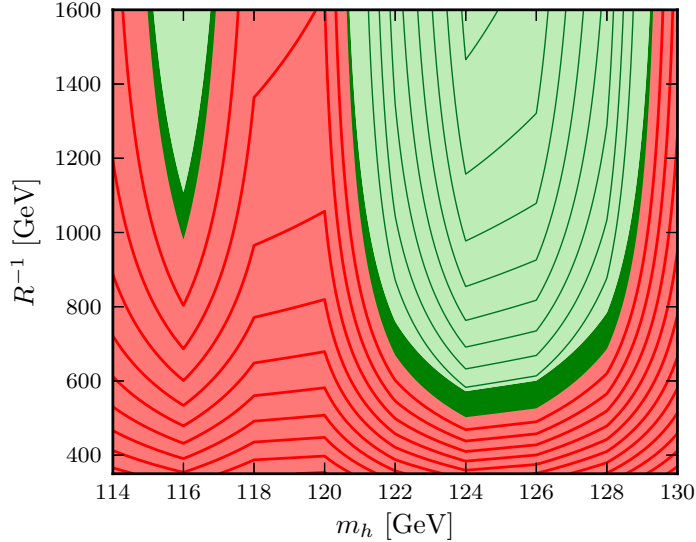}%
\end{center}
\caption{Exclusion plot for MUED showing the region excluded at a 95~\% confidence in red (medium grey) and the allowed region in light green (light grey). The strip in dark green (dark grey) is the extra region allowed when loop-corrected KK masses are used instead of tree-level masses.}
\label{fig:muedexc}
\end{figure}

To test MUED, we compared the predicted signal enhancement with the experimental values in the $H\to\gamma\gamma$, $h\to ZZ\to4\ell$ and $h\to WW\to 2\ell 2\nu$ channels. The predicted values are shown in Fig.~\ref{fig:muedmu}
for varying $R^{-1}$ and $m_H$. Under some approximations we constructed a likelihood function for data, given our model (for full details see Ref.~\cite{Belanger:2012mc}); a 95~\% confidence level exclusion plot based on the July 2012 ATLAS \cite{Aad:2012tfa} and CMS \cite{Chatrchyan:2012ufa, CMS-PAS-HIG-12-015, CMS-PAS-HIG-12-016} is shown in Fig.~\ref{fig:muedexc}.
This shows that the Higgs data exclude $R^{-1}$ below 500~GeV.

\section{4D Composite Higgs Model}

The work in this section was done by D.~Barducci, A.~Belyaev, M.S.~Brown, S.~de Curtis, S.~Moretti, and G.M.~Pruna; see Ref.~\cite{Barducci:2013wjc} for details. 
The 4D Composite Higgs Model (4DCHM) was proposed in Ref.~\cite{DeCurtis:2011yx}. 
In this model, the Higgs is posited to be a bound state of a sector of strongly interacting fermions. It is a pseudo Nambu-Goldstone boson so it can be as light as 125~GeV. 
There are other fermionic and bosonic resonances from the strong sector ($W^\prime$ and $Z^\prime$, gauge bosons and $t^\prime$ and $b^\prime$ quarks) that can contribute via loops to $\kappa_g$ and $\kappa_\gamma$. 
The tree-level Higgs couplings are modified from the SM values due to mixing of the new ``primed'' particles with their SM equivalents. We considered the full effects of this mixing and of the new particles running in loops for the first time, in contrast to previous work \cite{Falkowski:2007hz, Azatov:2011qy, Azatov:2012rd, Gillioz:2012se, Azatov:2012ga}.

\begin{figure}[tb]
\begin{center}
\includegraphics[width=0.5\linewidth]{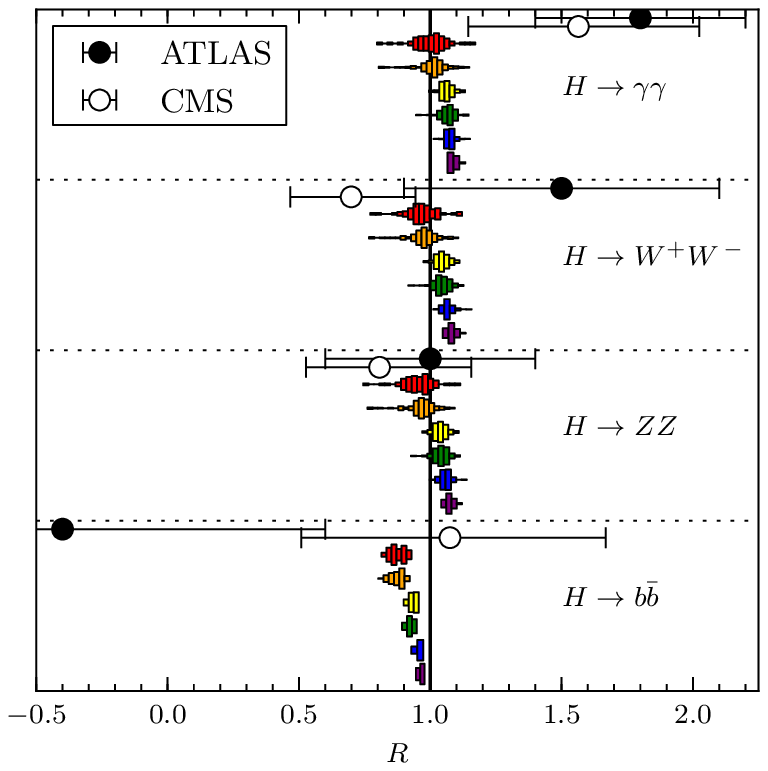}%
\includegraphics[width=0.5\linewidth]{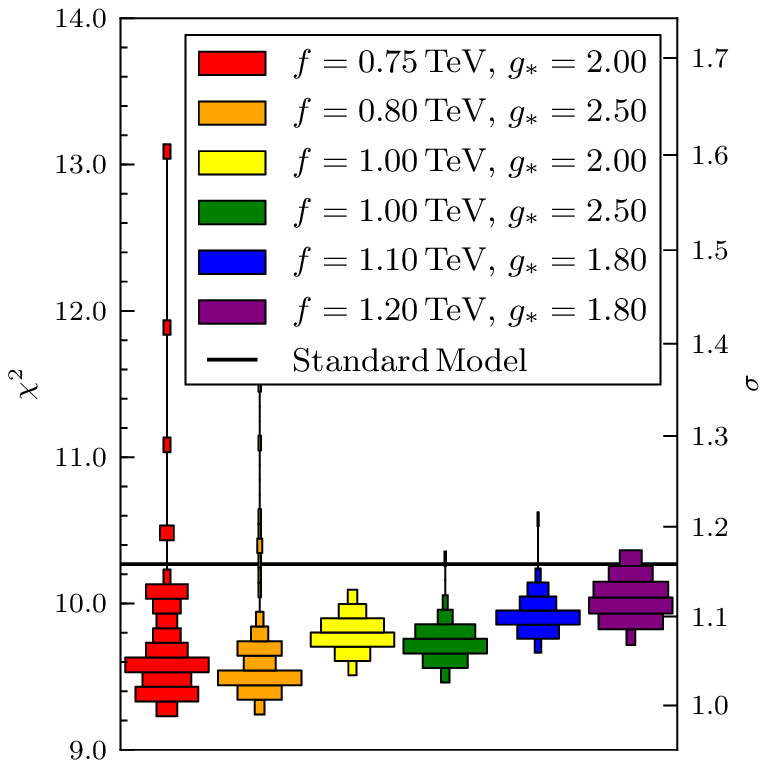}
\caption{Left: Histogram showing relative number of points in the random scan that yielded particular expected values of $\mu_X$ for four final states $X=\gamma\gamma, WW, ZZ, b\bar{b}$: the experimental values are plotted for comparison. The scan was performed for several benchmark values of $f$ and $g^*$ (shown as separate histograms) which are in the same order, top to bottom, as shown in the legend in the right-hand figure.
Right: Histograms showing number of points in scan that yielded particular $\chi^2$ values relative to experiment. As previously, the benchmarks are shown in the same order (left to right) as in the legend (top to bottom).}
\label{fig:4dchm}
\end{center}
\end{figure}

The parameter space for 4DCHM is much bigger than that of MUED. 
There is a compositeness scale $f$ and a new gauge coupling $g^*$. 
New fermionic resonances can mix with each other and the SM fermions, introducing a further nine parameters. 
These parameters give rise to the SM top and bottom quark masses, the Higgs mass and the masses of the new resonances. 
We chose not to invert the relations to make $m_t,m_b,m_H$ input parameters. 
Instead, we fixed $f$ and $g^*$ to a selection of benchmark points and performed a random scan over the remaining parameters, rejecting points that lead to unphysical masses, or low masses of $W^\prime,Z^\prime,t^\prime,b^\prime$ that would fall foul of direct detection and electroweak precision tests. 
For each surviving point in the scan we calculated the $\mu_X$ parameters for $X=\gamma\gamma,WW,ZZ,bb$. 
Fig.~\ref{fig:4dchm} (left) 
shows histograms for each $X$ and choice of $(f,g^*)$; larger bars correspond to values of $\mu_X$ that were more common in the scan. 
The ATLAS and CMS measured values are also plotted: we used the CMS~\cite{CMS-PAS-HIG-12-045} and ATLAS~\cite{ATLAS-CONF-2012-170} data released for the Hadron Collider Physics Symposium in December 2012.

The goodness of fit to the ATLAS and CMS data can be expressed by computing
\begin{equation}
    \chi^2 = \sum_{i\in\{\text{ATLAS},\text{CMS}\}} \sum_{X} \frac{(\mu_X - \mu_{X,i})^2}{\sigma_{X,i}^2}
\end{equation}
where $\mu_{X,i}$ is the ATLAS or CMS measured value and $\sigma_{X,i}$ is the standard error on the measurement. Fig.~\ref{fig:4dchm} (right) is a histogram that shows the relative popularity of different $\chi^2$ values in the scan. Most points in the scan have a \emph{lower} $\chi^2$ value than the SM value, showing that the Higgs data do not at present constitute any additional challenge to the validity of the 4DCHM. All points lie within two standard deviations of the data.

\section{Conclusions}
We have shown how one can use the recent LHC Higgs mass and couplings measurements to constrain BSM theories. For the example of MUED we were able to exclude the compactification scale to $R^{-1} > 500$~GeV. For the 4DCHM there were many more parameters and so a scan was performed. After rejecting scan points that contradicted other experimental tests, we found that all remaining points were easily compatible with the current Higgs data.

\bibliographystyle{JHEP}
\bibliography{brown}

\end{document}